\documentstyle[prb,aps,epsf]{revtex}

\def\nuc#1#2{\relax\ifmmode{{}^{#1}{\relax\text{#2}}}\else{${}^{#1}$#2}\fi}
\def\nuc#1#2{${}^{#1}$#2}

\begin{document}
\draft

\title{Phase Diagram of the Superfluid Phases of $^{3}$He in 98\% Aerogel}

\author{G. Gervais$^{1}$, K. Yawata$^{1}$, N. Mulders$^{2}$ and W.P. Halperin$^{1}$}

\address{
$^{1}$Department of Physics and Astronomy, Northwestern University, Evanston, Illinois, 60208 USA\\
$^{2}$Department of Physics and Astronomy, University of Delaware, Newark, Delaware, 19716 USA\\}

\date{\today}

\twocolumn[\hsize\textwidth\columnwidth\hsize\csname@twocolumnfalse\endcsname

\maketitle
\begin{abstract} The phase diagram of the superfluid phases of $^{3}$He in 98\% aerogel was determined in the range of pressure
from 15 to 33 bars and for fields up to 3 kG using high-frequency sound. The superfluid transition  in aerogel at 33.4 bars is
field independent from 0 to 5 kG and shows no evidence of an
$A_{1}-A_{2}$ splitting.  The first-order transition between the A and B-phases is  suppressed by a magnetic field, and
exhibits strong supercooling at high pressures. We show that the  equilibrium phase in zero applied field  is the B-phase with
at most a region of A-phase {\small $\alt$} 20 $\mu$K just below T${_c}$ at a pressure of 33.4 bars.  This is in contrast to
pure
$^{3}$He which has a large stable region of A-phase and a polycritical point.  The  quadratic coefficient for  magnetic field
suppression  of the AB-transition,
$g_{a}(\beta)$,  was obtained. The pressure dependence of 
$g_{a}(\beta)$ is markedly different from that for the pure superfluid, $g_{0}(\beta)$,   which diverges at a polycritical
pressure of 21 bars.    We compare our results with calculations from the homogeneous scattering model for $g_{a}(\beta)$,
defined in a Ginzburg-Landau theory in terms of strong-coupling parameters $\beta$.  We 
 find qualitatively good agreement with the experiment if the strong-coupling corrections are rescaled from known
values of the $\beta$'s for pure
$^{3}$He, reduced by the suppression of the superfluid transition temperature.  The calculations indicate that the polycritical
pressure in the aerogel system is displaced well above the melting pressure and out of experimental reach. We cannot
account for the puzzling supercooling of the aerogel AB-transition in zero applied field within the framework of known
nucleation scenarios.
\end{abstract}
 ]

\narrowtext

\section{Introduction} Quenched disorder in condensed matter systems is manifest in a wide variety of materials
 from glassy solids and liquid crystals to the mixed state of superconductors.  It  arises in diverse phenomena ranging from
cosmological models for the evolution of the universe to vortex tangles in superfluid $^{4}$He.  Disorder in superfluid
$^{3}$He is of special interest since the order parameter structure of this superfluid is non-trivial, although well known, and
its various phases exhibit a number of spontaneously broken symmetries.  Quenched disorder in a superfluid can be generated
extrinsically by a random impurity field with inhomogeneity on a length scale short compared to the coherence length.  In the
present case this is achieved by imbibing $^{3}$He into silica aerogel, a highly-porous material made of randomly
inter-connected strands of SiO$_{2}$.  

Aerogels have been used to study liquid crystals\cite{Wu92}, superfluid
$^{4}$He\cite{Chan1}, $^{3}$He-$^{4}$He mixtures \cite{Chan2} and superfluid $^{3}$He\cite{Porto95,Spra95}.  Glassy effects
have been observed in the liquid crystal-aerogel systems.  In $^{3}$He-$^{4}$He mixtures a profound influence of aerogel on the
phase diagram\cite{Chan2} was reported. Superfluid
$^{3}$He in aerogel was found to have a suppressed, but  relatively sharp, transition temperature and the order parameter
appears to be reduced\cite{Porto95,Spra95}. However, the nature of the phase diagram, and identification of the
thermodynamically stable phases, have not yet been clearly established.  Here we report the use of high resolution transverse
acoustic impedance to map out this phase diagram.

The $^{3}$He-A and -B superfluid phases were discovered in 1972 by Osheroff, Richardson and Lee\cite{Osh72}. The order parameter is
now established to be a {\it p}-wave spin-triplet which has two thermodynamically stable superfluid phases in zero field. The
A-phase is the axial state which separately breaks spin and orbit rotational symmetries.  The B-phase is the isotropic state
which breaks relative spin-orbit symmetry.  The stability of the A-phase over the B-phase at elevated pressure, above the
polycritical pressure of 21 bars in zero field, is a consequence of strong coupling in the quasiparticle interactions.  After
30 years of extensive experimental and theoretical investigation
$^{3}$He is the best understood of all unconventional superfluids or superconductors.  Motivation to investigate disorder in
this superfluid derives in part from our need to understand  impurity effects in this, and in similar, unconventionally paired
systems.   New families of superconducting materials such as Sr$_{2}$RuO$_{4}$\cite{Ruthenate}, URhGe\cite{Flouquet01}, and
organic conductors \cite{Organic} may be unconventional superconductors, and in some cases, a {\it p}-wave structure has been
suggested. 

At millikelvin temperatures $^{3}$He is the purest material in nature.  Its properties have been investigated extensively as a
system entirely free of impurity scattering other than at surfaces.  The influence of aerogel on $^{3}$He is to suppress
the transition retaining a narrow width and to alter the behavior of the superfluid phases. In
contrast, for surface scattering, the orientation and the amplitude of the
order parameter are both constrained and the superfluid becomes spatially inhomogeneous.  The use of highly-porous aerogels to introduce
impurity scattering  in
$^{3}$He has provided a significant opportunity to learn about unconventional pairing states. The first observations of
superfluidity were torsional oscillator experiments to measure the superfluid density performed at Cornell \cite{Porto95} and
Nuclear Magnetic Resonance (NMR) at Northwestern \cite{Spra95,Spra96}. These results were found to be consistent with
theoretical models for impurity scattering
\cite{Thune98}. In its simplest form, the Homogeneous Scattering Model (HSM), the suppression of the superfluid transition  is
given by the well-known Abrikosov-Gorkov formula

\begin{equation}
ln(\frac{T_{c0}}{T_{ca}})=\Psi(\frac{1}{2}+\frac{1}{2}\frac{\xi_{0}}{l_{tr}}\frac{T_{c0}}{T_{ca}})-\Psi(\frac{1}{2})
\label{AGformula}
\end{equation}
where $T_{ca}$($T_{c0}$) is the superfluid transition temperature of the aerogel (bulk) system, $\xi_{0}\equiv \hbar v_{f}/2\pi
k_{B}T_{c0}$ is the bulk coherence length, and $l_{tr}$ is the transport mean free path. These earlier results triggered numerous
experimental investigations\cite{Alles98,Golov99,Matsubara00,Barker00,Bunkov00,Lawes00} and theoretical
work\cite{Thune98,Volo96,Bara96,Thuneberg,Hanninen98,Einzel98,Higa99,Priya01,Mineev02} aimed at characterizing and understanding the
properties of the `dirty' superfluid formed inside the aerogel matrix. However, the identification of the superfluid state in the
$^{3}$He-aerogel system has been controversial, and only recently has there been agreement as to which are the superfluid phases
observed.

The first NMR
measurement in pure
$^{3}$He and 1.2 kG suggested a superfluid in an equal-spin pairing (ESP) state\cite{Spra95} similar to the bulk A-phase.
With $^{4}$He additions  a non-ESP state was found\cite{Spra96} like the bulk B-phase.   NMR measurements at lower fields ($\sim$50 G)
without $^{4}$He found evidence for a B-phase superfluid in aerogel \cite{Alles98} and  Barker {\it et al.}\cite{Barker00} found a
transition between a ESP and a non-ESP state  at 284 G with
$^{4}$He coverage. This transition was found to supercool quite readily and  was identified to take place between A and B
superfluids, but it should be kept in mind that the orbital symmetry of the order parameter  in aerogel has not yet been
determined.  The aerogel AB-transition was also observed recently by a vibrating viscometer at low pressures
\cite{Pickett01}, and was studied near the bulk polycritical pressure (PCP)  using high-frequency sound\cite{Gervais01}. The reason for
covering the strands with some $^{4}$He (2 or more atomic layers) is to replace magnetic solid $^{3}$He that contributes to
measurement of the $^{3}$He magnetization and may also affect the nature of the scattering and possibly properties of the dirty
superfluid.   In the present work, we present a comprehensive acoustics study of the phase diagram of the A and B-superfluid
phases  of
$^{3}$He in a 98\% aerogel without $^{4}$He and from 0 to 3 kG.

\section{Experimental} High-frequency sound ($\sim$MHz) has proven to be a powerful tool to study the properties of pure $^{3}$He in
the normal Fermi liquid and in the superfluid state. For example, Landau's seminal prediction of collisionless sound,
 called zero-sound, in a Fermi liquid was experimentally verified with high-frequency sound attenuation
measurements\cite{Abel66}.  High-frequency sound also couples strongly to the order parameter collective modes (OPCM) of
the superfluid and  numerous OPCM have been observed in both the A and B-phases of pure $^{3}$He (see
\cite{Halp90} for a review). Transverse sound was predicted to exist in normal $^{3}$He by Landau, and was shown to propagate
in $^{3}$He-B from the observation of an acoustic Faraday effect\cite{Lee99}. In what follows, we describe a technique used to
probe the phase diagram of $^{3}$He in aerogel using both transverse and longitudinal high-frequency sound waves.  

\begin{figure}[t]
  \begin{center}
   \leavevmode
    \hbox{\epsfxsize=.8\columnwidth{\epsfbox{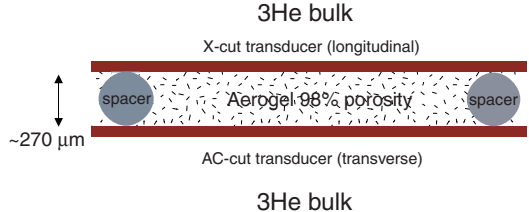}}}
  \end{center}
  \caption{Schematics of the acoustic cavity. The X-cut (longitudinal sound)  and AC-cut (transverse sound) transducers are
separated by 270
$\mu$m spacers and the 98\% porous aerogel was grown {\it in situ}. Each transducer has two active sides; one probes the
aerogel-filled cavity while the other probes the bulk liquid outside the cavity.}
  \label{cell}
\end{figure}

The acoustic technique is similar to that reported previously\cite{Nomura00,Roach76} and a schematic of our acoustic cavity is
depicted in Fig.\ref{cell}. The cavity was formed with two quartz transducers separated by two parallel stainless steel
 wires of diameter
$d=270$ $\mu m$\cite{Gervais00}.  One transducer was AC-cut for transverse sound, and the other X-cut for longitudinal sound,
with a diameter of 9.5 mm.  Their fundamental frequencies were 4.8 MHz and 2.9 MHz, respectively. This arrangement  allowed us
to perform experiments with either transverse or longitudinal sound.  The aerogel was grown {\it in situ},  in the volume
between the transducers. Each transducer has two active sides; one probes the aerogel-filled cavity, and the other the bulk
liquid outside the cavity. All experiments were performed with pure $^{3}$He which was verified to contain less than 250 ppm of
$^{4}$He, much less than the amount required to cover the aerogel strands with one atomic layer of $^{4}$He. 

\begin{figure}[t]
  \begin{center}
   \leavevmode
    \hbox{\epsfxsize=1.\columnwidth{\epsfbox{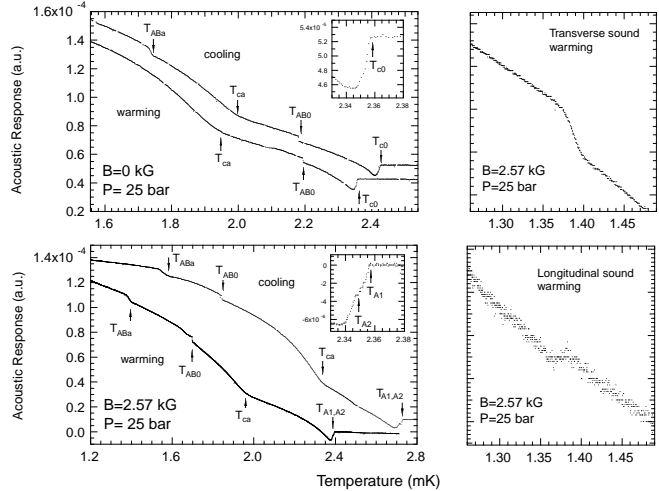}}}
  \end{center}
  \caption{Left: transverse acoustic response at 25 bars and zero applied field (upper panel) and 2.57 kG (lower panel). In each
panel, the data in the upper (lower) trace is taken on cooling (warming). The temperature scale is for warming only; for cooling
the traces were offset for clarity. The various transitions in the bulk and aerogel superfluids are indicated by arrows. The
inset is an enlarged view of the bulk superfluid transition
$T_{c0}$ in zero  field (upper), and 2.57 kG (lower) which shows the bulk $A_{1}$ and $A_{2}$ transitions. Right: an enlarged
view of the aerogel AB-transition on warming is shown at 25 bars and 2.57 kG for the acoustic response of transverse
sound at 8.691 MHz in the upper panel and longitudinal sound at 14.635 MHz in the lower panel.  }
\label{trace}
\end{figure}

The electrical impedance of the transducers was measured using a continuous wave spectrometer. The measurements were performed
at a fixed frequency corresponding to odd harmonics of the fundamental resonance, with a frequency modulation of 400 Hz and
modulation amplitude of 3 kHz.  In the case of longitudinal sound, the medium  inside the cavity is of sufficiently low
attenuation that a standing wave pattern is established throughout the cavity.   Small changes in the attenuation and velocity,
induced by changes in temperature or pressure, produced changes in the electrical impedance of the transducers that can be
detected by the spectrometer
\cite{Nomura00,Gervais00}.   For transverse sound, the highly attenuating medium prohibits the existence of well-defined
standing waves and the measurement is similar to that of an acoustic impedance measurement\cite{Roach76}. It is not possible
with this technique to separate  individual contributions from attenuation, sound velocity, or coupling to collectives modes in
the transverse acoustic impedance.  However, we have found that,  at low frequencies ($\alt$ 10 MHz), the transverse sound
impedance changes abruptly at all of the known phase transitions  in each of the bulk and aerogel superfluids, Fig.\ref{trace}.

The transverse acoustic response at 8.691 MHz and a pressure of 25 bars is shown in the left panels of Fig.\ref{trace} on
cooling  (upper trace) and  warming (lower trace) and at zero (upper panel) and 2.57 kG applied magnetic field (lower panel).
The phase transitions  in the bulk liquid and the aerogel are denoted by arrows.  On cooling, and in zero applied field, we
successively observed the bulk superfluid ($T_{c0}$),  bulk AB ($T_{AB0}$), aerogel superfluid ($T_{ca}$) and aerogel AB
($T_{ABa}$) transitions.  The thermometry scales for cooling and warming are different.  The scale shown in the figure is for
slow warming such that equilibrium is assured between the LCMN thermometer used in low field, the melting curve thermometer,
and the aerogel sample. For the results in Fig.\ref{trace} and Fig.\ref{trace2}, we corrected the thermometry for the more 
rapid cooling experiments using the temperature dependence of the acoustic impedance established in equilibrium during
warming.  This provides a convenient and intrinsic secondary thermometer, but for clarity the cooling traces are offset.  When
a magnetic field is applied, we also observed the bulk A$_{1}-A_{2}$ splitting as a `knee'  in the acoustic response,  as 
shown in the inset of the lower left panel. The bulk
$A_{1}$ and
$A_{2}$-transitions are completely resolved  in the acoustic trace at a field of $\sim$ 5 kG and above.

In the right panels of Fig.\ref{trace} we show a direct comparison of the signatures from transverse and longitudinal sound for
the aerogel AB-transition observed on warming at 25 bars and 2.57 kG.  A small jump  in the acoustic trace of longitudinal sound,
coincident in temperature with that of transverse sound, was observed when the AB-transition occured at temperatures sufficiently well
below
$T_{ca}$.  The condition for observing the signature in longitudinal sound was that the transition be either supercooled or
that it appear on warming in a field
$B\agt$ 2 kG.  However, we cannot determine whether this jump in the longitudinal sound trace
 arises  from a change in attenuation due to collective modes, or from quasiparticle excitations.  The observation of the aerogel
AB-transition in longitudinal sound ensures that the transition observed  with transverse sound is not a local effect occurring
near the surface of the transducer.  We have shown previously\cite{Nomura00,Gervais00} that a well-defined longitudinal
sound mode can be established in our acoustic cavity, and therefore the AB-transition observed with this mode reflects the behavior
of the superfluid over the entire aerogel sample. We have also verified that the change of slope in the transverse acoustic
trace labeled $T_{ca}$ corresponds to the temperature at which the attenuation of longitudinal sound decreases at the onset of
the superfluid transition\cite{Nomura00}.   However, the wider range of observability of the aerogel AB-transition by
transverse sound, as compared with longitudinal sound, and its higher precision make it a better tool to map out the phase
diagram  of $^{3}$He in 98\% aerogel. The frequency dependence of the bulk superfluid transition $T_{c0}$, as observed with
transverse sound, was also systematically studied at a pressure of 17 bars and for frequencies ranging from 3 MHz to 55 MHz.  We
found that the transverse acoustic signature signaling $T_{c0}$ is weakly frequency dependent but that it  recovers the
transition temperature
$T_{c0}$ in the low-frequency limit, $\alt$10 MHz.  The transition temperatures observed in our sample are in excellent
agreement  with those reported elsewhere 
\cite{Spra95,Barker00,Matsu97} for the same density of aerogel.  In what follows, the (P,T,B)  phase diagram of
$^{3}$He in aerogel was determined using transverse sound at a frequency of 8.691 MHz.

\section{Results and discussion}

\subsection{P-T-B dependence of the AB-transition in aerogel}

One of the key issues in the study of superfluid $^{3}$He in aerogel is to determine in what way the superfluid phase diagram
is modified by impurity scattering. Unlike {\it s}-wave superconductors for which only magnetic impurity scattering is pair breaking,
{\it p}-wave Cooper pairs are sensitive to all forms of scattering.  For superfluid $^{3}$He in aerogel, there is now general
agreement that impurity scattering suppresses the superfluid transition temperature and the amplitude  of the order parameter,
and that a non-ESP phase similar to the bulk B-phase is favored near zero field. However, most experiments have  been performed
in different regions of P, T, and B and also under different experimental conditions, {\it e.g.} with or without
$^{4}$He preplating on the aerogel strands and with different aerogel densities.  In particular, the phase diagram of the pure
superfluid phases of pure $^{3}$He in  98\% aerogel has not been established; the zero-field phase diagram is largely unknown.
We emphasize that only the  spin structure of the pairing state of
$^{3}$He in aerogel has been identified through NMR measurement of the magnetization\cite{Spra96,Barker00} or the stability of
the transition in a magnetic field
\cite{Pickett01,Gervais01}.   In the latter case, assuming triplet pairing, we can infer the spin component of the order
parameter (ESP or non-ESP) from the field dependence of the phase boundaries. The orbital symmetry is more
elusive. For example, it would be helpful to investigate the OPCM  which couple to orbital degrees of freedom of the order
parameter, but they have not yet been observed
\cite{Nomura00}.   Nonetheless, we use the earlier notation\cite{Barker00,Pickett01,Gervais01} `A' and `B'  for the aerogel
superfluid phases corresponding to ESP and non-ESP states, by analogy with the bulk, and we discuss later implications from our
measurements for the orbital symmetry of the aerogel superfluid.   

\begin{figure}[b]
  \begin{center}
   \leavevmode
    \hbox{\epsfxsize=1.\columnwidth{\epsfbox{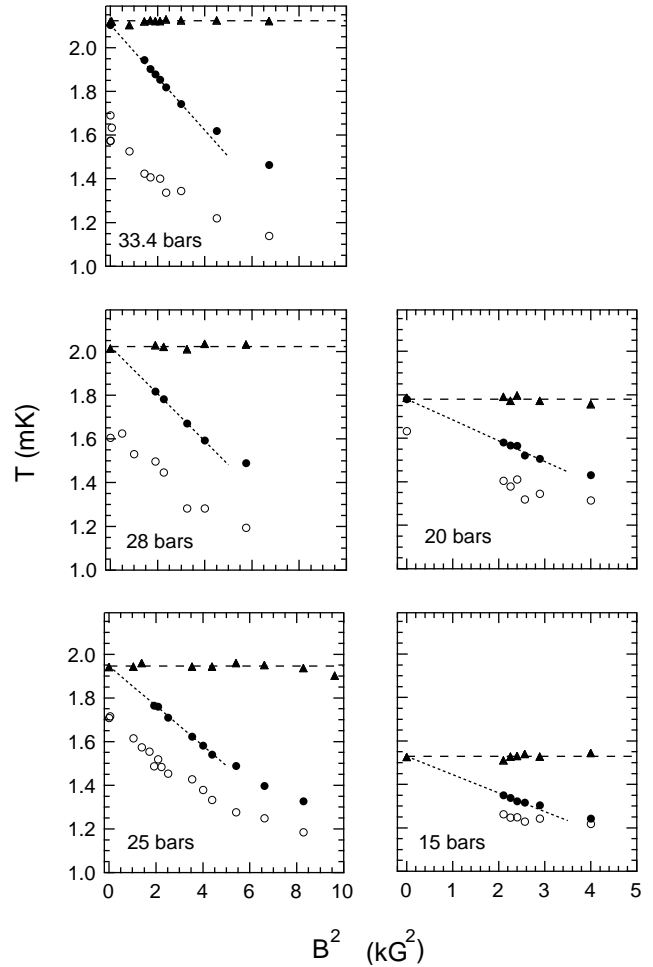}}}
  \end{center}
  \caption{Phase diagrams of the superfluid phases in aerogel at various pressures in a magnetic field. The triangles are the
superfluid transition temperature, $T_{ca}$, determined from transverse acoustics. The supercooled aerogel AB-transitions,
$T_{ABa}$,  are shown  on cooling (empty circles) and in equilibrium on warming (filled circles). The data are plotted as
function of
$B^{2}$ to illustrate the quadratic suppression of the AB-transition at low field. The dashed lines show the average value of
$T_{ca}$ and the dotted lines are an extrapolation of $T_{ABa}$  to zero field. Note that the field axis is different for the left
and right panels. } 
  \label{diagrams}
\end{figure}

The phase diagrams of the aerogel AB-transition as a function of $B^2$ and for various pressures are shown in
Fig.\ref{diagrams}. The triangles denote the superfluid transitions, $T_{ca}$, as determined by the change of slope in the
transverse acoustic trace, the filled circles are the equilibrium AB-transitions taken on slow warming, and the empty circles
are the supercooled AB-transitions. The long-dashed lines show the average values of $T_{ca}$ and the dotted lines are
extrapolations to zero field.  With our technique, we were unable to observe directly the aerogel AB-transitions on warming at
fields below 1.4 kG, i.e. for
$T_{ABa}/T_{ca}\agt 0.9$. An independent check on the validity of our extrapolation  to zero field will be discussed below. 

The magnetic field independent transition from normal to superfluid state, and field-dependent transitions from A to
B-superfluids strongly suggest that these are transition between normal fluid to ESP superfluid states, and between ESP and
non-ESP states, as in the bulk.  The strong supercooling of the AB-transition, even in zero applied field, shows that the
transition is first-order. However, there are key differences in the aerogel phase diagram as compared to the bulk. In
particular, the zero-field  equilibrium region of aerogel A-phase is extremely narrow,$\alt$ 20
$\mu$K at 33.4 bars, if it exists at all.

In order to locate the AB-phase boundary at 33.4 bars in zero applied field, we have performed a series of temperature sweeps
in which the temperature was slowly raised from  low-temperature ($\sim$ 0.6 mK) to a temperature in the  vicinity, but
slightly less than, $T_{ca}$, and then rapidly cooled. If the AB-phase boundary were reached on warming, the supercooled
AB-transition would be observed on cooling, whereas in the opposite case  the aerogel superfluid would remain in the B-phase
with no evidence for an AB-transition.  This procedure depends on the absence of what is called a memory effect for secondary
nucleation which we verified indpendently in modest magnetic fields\cite{Gervais02}. 

\begin{figure}[t]
  \begin{center}
   \leavevmode
    \hbox{\epsfxsize.9\columnwidth{\epsfbox{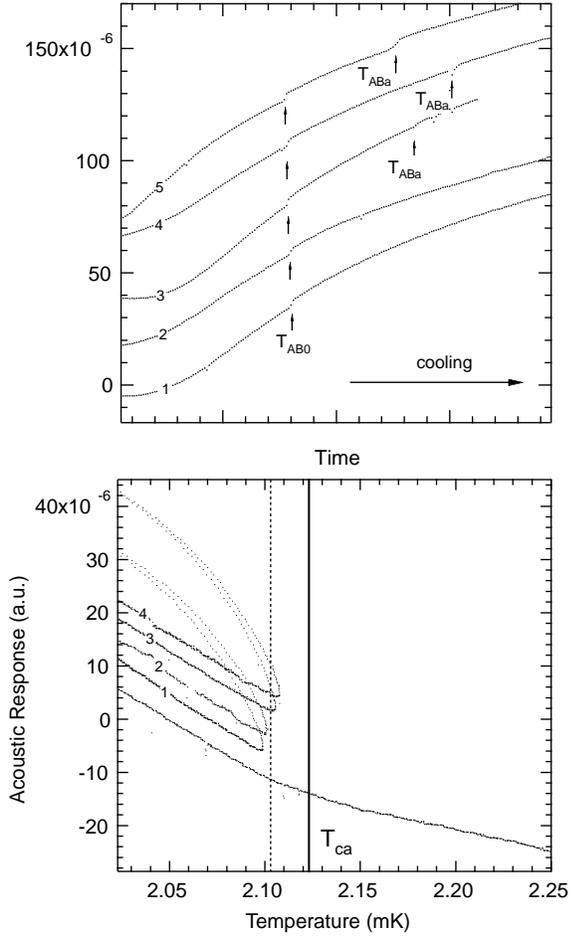}}}
  \end{center}
  \caption{Lower panel: transverse acoustic response at 33.4 bars and zero applied field during slow warming in the vicinity
of $T_{ca}$ and subsequent rapid cooling. The traces labeled 1 to 4 correspond to several maximum temperatures reached prior to rapid
cooling. The corresponding cooling traces are shown in the upper panel as a function of time and the bulk and aerogel
AB-transitions are denoted by arrows. For traces 1 and 2 the aerogel AB transition was not crossed on warming and for traces 3
and 4 it was.  The trace labeled 5 shows the aerogel AB-transition but for a cooling  experiment originating from the normal
state. In the lower panel, the vertical solid line is
$T_{ca}$ as determined from the change in slope of the acoustic response on warming, and the dotted line is a lower bound on
the  temperature for the aerogel AB-phase boundary, $\sim$20
$\mu$K below $T_{ca}$.}
  \label{trace2}
\end{figure}

 Fig.\ref{trace2} shows these various traces upon warming to a temperature in the vicinity of
$T_{ca}$ (lower panel and traces labeled 1 to 4, vertically offset for clarity), and  then rapidly cooled.  The thermal
disequilibrium during sudden cooling is sufficient to make the cooling traces appear to move up in this figure. The lowest trace
shows the complete acoustic record with $T_{ca}$ indicated by a change of slope and marked by a solid vertical line. 

In the
upper panel,  we show the same acoustic traces for rapid cooling (1 to 4) as a function of time  while the trace labeled 5 is
for a cooling experiment originating from the normal state. We do not observe the aerogel AB-transition upon cooling in the
trace 1 and 2 (and also for any partial warmup to temperatures below those shown here), while in the traces 3 and 4 the aerogel
AB-transition is observed on cooling.  This shows that the equilibrium aerogel AB-transition  lies at a temperature somewhere
between the traces 2 and 3, which is indicated in the lower panel of Fig.\ref{trace2} by a dotted line.  This is close to
$T_{ca}$ ($\alt$20 $\mu$K) and is approximately the  width of $T_{ca}$ itself ($\sim$30 $\mu$K), sufficiently close that 
 we cannot say that the normal state of the $^{3}$He fluid in aerogel had not been reached somewhere in the sample. 
Consequently, we have inferred that the  region of A-phase in aerogel in zero applied field is extremely narrow, $\alt$20
$\mu$K at a pressure of 33.4 bars. We have also verified at a pressure of 20 bars that the AB-phase boundary in zero applied
field is indistinguishable from
$T_{ca}$ using the same slow warming and quench-cooling method.  
 
\begin{figure}[t]
  \begin{center}
   \leavevmode
    \hbox{\epsfxsize=1.\columnwidth{\epsfbox{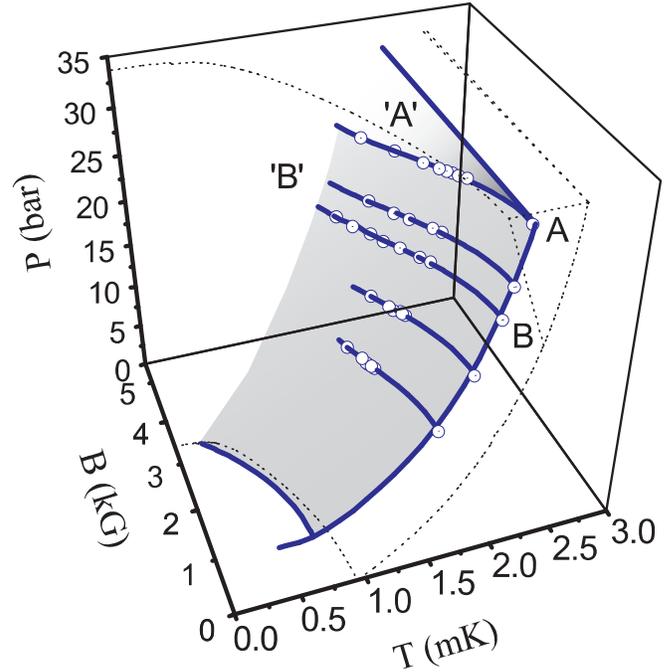}}}
  \end{center}
  \caption{Three-dimensional phase diagram (P,T,B) of the A and B superfluid phases of $^{3}$He in 98\% aerogel. The aerogel
phases are labeled `A' and `B' and are delineated by solid lines while for the pure phases they are labeled A and B and shown with
dotted lines. The shaded area shows the volume spanned by the equilibrium A-phase in aerogel. The open circles are data from
the present work, and the lines connecting them are fits to the data. At pressures below 10 bars, $T_{ca}$, was taken from
Matsumoto {\it et al.}\protect\cite{Matsu97}, together with the field dependence of the AB-transition observed by the Lancaster
group at 4.8 bars\protect\cite{Pickett01}. } 
  \label{3d}
\end{figure}

From the data of Fig.\ref{diagrams} we construct a three-dimensional phase diagram for pressure, temperature and magnetic field
for the superfluid phases of pure $^{3}$He in 98\% aerogel, Fig.\ref{3d}. The relative stability of the superfluid phases of
the pure and dirty superfluids  can be directly visualized and compared. The data of the equilibrium AB-transition  from
Fig.\ref{diagrams} are shown as empty circles, and the thick solid lines are a smoothed fit of  $T_{ABa}$ and
$T_{ca}$. Below 10 bars, the data from Matsumoto {\it et al.} \cite{Matsu97} were used to describe $T_{ca}$.  They found a
critical pressure of $\sim$ 6 bars in the zero-temperature limit. The thick solid curve at a pressure of 4.8 bars was taken
from the field dependence of the aerogel AB-transition measured by the Lancaster group with a vibrating
viscometer\cite{Pickett01}, but with its superfluid  transition adjusted to match that of Matsumoto {\it et al.}\cite{Matsu97}. 
Above 10 bars our data for
$T_{ca}$ in zero field are similar from that of Matsumoto {\it et al.}\cite{Matsu97} and the data points are not
shown for reasons of clarity. The aerogel phases are denoted by `A' and `B' while the pure phases are denoted by A and B. The
shaded  volume emphasizes the aerogel A-phase opening up with applied magnetic field, and the dotted lines are the pure phase
diagram shown here for comparison\cite{Hahn95}.  The field dependence  of the aerogel superfluid transitions
$T_{ca}$ plotted as a solid line at 33.4 bars will be discussed later. 

The three-dimensional plot of the (P,T,B) phase diagram of $^{3}$He in 98\% aerogel shows the effects of impurity scattering on
the equilibrium A-phase. The equilibrium region of A-phase is destabilized by impurity scattering in zero field;
however, we believe that there is a very thin sliver of equilibrium A-phase that gives rise to supercooling of the  A-phase
observed in zero applied field\cite{Gervais01} at pressures above 15 bars.  Barker {\it et al.}\cite{Barker00} observed a
metastable aerogel AB-transition on cooling using an NMR technique at 284 G and with $^{4}$He preplating. At 32 bars, an
equilibrium  region of A-phase of 70 $\mu$K was inferred from the data. This is somewhat larger than our results at the same
field.  The magnetization discontinuity at the AB-transition known to occur in the bulk system was not observed in
aerogel,  making it difficult to locate the equilibrium AB-transition.  Furthermore, given the scatter of  the
data\cite{Barker00} and the thermometry resolution $\sim20$ $\mu K$, it seems plausible  that the A-phase region might have
been smaller than 70 $\mu$K and consistent with our findings here.

\subsection{Magnetic suppression of the AB-transition and the Homogeneous Scattering Model}

The Ginzburg-Landau (GL) theory for superfluid $^{3}$He describes the free energy near the transition temperature expanded in
powers of the order parameter.  With this approach the relative stability of various possible {\it p}-wave states can
be explored in terms of the expansion coefficients of the theory\cite{Mermin}. An extension to the dirty
superfluid has been developed\cite{Thune98} using a model that describes elastic quasiparticle scattering. There are five possible
fourth order combinations of the order parameter that are invariant under all the symmetries of the {\it p}-wave superfluid.
These fourth-order terms are characterized by the coefficients,
$(\beta_1,\beta_2,\beta_3,\beta_4,\beta_5)$,  which in the GL-theory are determined by  thermodynamic quantities such as heat
capacity, the magnetization, the phase diagram and the NMR frequency shift.  In principle one could uniquely determine the
$\beta_{i}$'s (thus the free energy functional) if five independent thermodynamic measurements were performed; the fact that in
bulk superfluid $^{3}$He there are only four such measurements is  unfortunate. Nonetheless, combinations of the $\beta_{i}$'s
can be extracted from experiment and are very helpful in the understanding the magnetic suppression of the  B-phase of
$^{3}$He in aerogel.

In the pure superfluid the GL-theory can be used to describe the suppression of the AB-transition by magnetic field only for
pressures less than 21 bars, the pressure of the polycritical point.  For superfluid $^{3}$He in aerogel, the data
from Fig.\ref{diagrams} suggest that this theory can be
applied at all pressures with 

\begin{equation} 1-\frac{T_{ABa}}{T_{ca}}=g_{a}(\beta)(\frac{B}{B_{0}})^2+{\cal O}(\frac{B}{B_{0}})^4,
\label{GL}
\end{equation} where $g_{a}(\beta)$ is a strong-coupling parameter defined in a manner similar to that of the pure
superfluid and $B_{0}$ is defined as,
\begin{equation} B_{0}=\sqrt\frac{8\pi^2}{7\zeta (3)}\frac{k_{B}T_{ca}}{\gamma\hbar}(1+F_{0}^{a}),
\label{fieldscale}
\end{equation} with $\gamma$ the gyromagnetic ratio of $^{3}$He and $F_{0}^{a}$ is a Fermi liquid parameter. In
the Homogeneous Scattering Model (HSM) which we describe below, $B_{0}$ is modified by impurity scattering according
to\cite{Priya01,Thuneberg}
\begin{equation}
B_{0}^{HSM}/B_{0}=\sqrt{\frac{7\zeta (3)[1-x\sum_{n=1}^{\infty}(n-1/2+x)^{-2}]}{\sum_{n=1}^{\infty}(n-1/2+x)^{-3}}}
\label{fieldscalehsm}
\end{equation} where $x\equiv\hbar v_{f}/4\pi k_{B}T_{ca}l_{tr}$ and $l_{tr}$ is the transport mean free path. This
correction is about 2.5\% at 25 bars with a 200 nm mean free path. We use the weak coupling 
approximation in Eq.\ref{fieldscale} which has been shown
to be appropriate for bulk $^{3}$He \cite{Tom}.  
Assuming that the aerogel AB-transition occurs between the axial and the isotropic states, as in the pure superfluid,  the coefficient
$g_{a}(\beta)$ is written as,

\begin{eqnarray} g_{a}(\beta)=\frac{\beta_{245}}{2(-3\beta_{13}+2\beta_{345})}\times\\
\left( 1+\sqrt{\frac{(3\beta_{12}+\beta_{345}) (2\beta_{13}-\beta_{345})}{\beta_{245}\beta_{345}}}\right)
\label{gtheory}
\end{eqnarray} 

where we have used the Mermin-Stare convention, $\beta_{ijk}\equiv\beta_{i}+\beta_{j}+\beta_{k}$.  In the weak-coupling limit,
$-\beta^{wc}_{5}=\beta^{wc}_{4}=\beta^{wc}_{3}=\beta^{wc}_{2}=-2\beta^{wc}_{1}=2\beta_{0}^{wc}$ with
$\beta_{0}^{wc}=7\zeta (3)N(0)/240(\pi k_{B}T_{c0})^2$ and N(0) the density of states. 
In the weak-coupling limit, $g_{a}(\beta^{wc})=1$.

\begin{figure}[t]
  \begin{center}
   \leavevmode
    \hbox{\epsfxsize=1.\columnwidth{\epsfbox{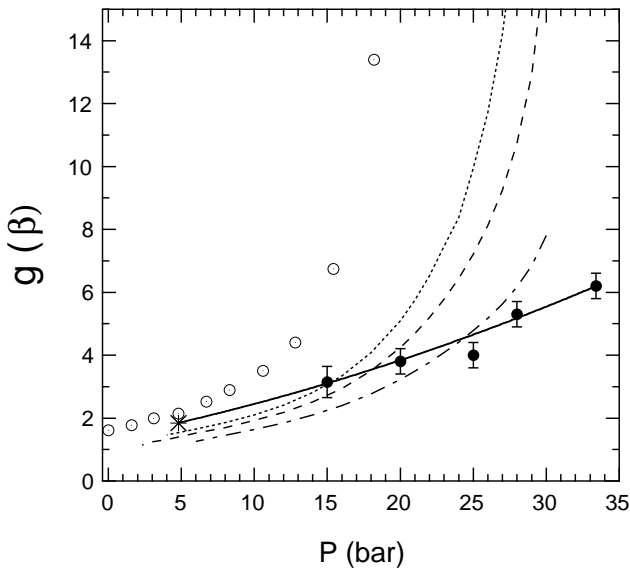}}}
  \end{center}
  \caption{Pressure dependence of the strong-coupling coefficient $g(\beta)$. The aerogel data from the present work  (solid
circles) are compared to the bulk (empty circles)\protect\cite{Hahn95}. The data at 4.8 bars denoted by a star are from Brussard
{\it et al.}\protect\cite{Pickett01}. The solid line is a guide-to-the-eye. The dotted line is the HSM with
$l_{tr}=200$ nm  and for which the pure $^{3}$He strong-coupling corrections are used. The same model was used with
strong-coupling corrections rescaled from pure $^{3}$He  by the factor
$T_{ca}/T_{c0}$ and two values of transport mean free path were choosen, $l_{tr}=200$ nm (dashed)
and 150 nm (dot-dashed).}
\label{gbeta}
\end{figure}

The coefficient $g_{a}(\beta)$  can be taken directly from the low-field slope, $m$, of the data in Fig.\ref{diagrams}, 
$m=-g_{a}(\beta)T_{ca}/B_{0}^{2}$. In Fig.\ref{diagrams}, the dotted lines show the quadratic suppression of $T_{ABa}$
for the smallest field at which the transition was observed, and the extrapolation to zero field was verified using the quench-cooling
method we have described above, see Fig.\ref{trace2}. The pressure dependence of $g_{a}(\beta)$ obtained from the data between
15 and 33.4 bars is shown in Fig.\ref{gbeta} (filled circles), together with the measured values from the bulk, $g_{0}(\beta)$,
taken from Hahn {\it et al.}.\cite{Hahn95} (open circles). For the aerogel data, $B_{0}$ was defined as in
the pure case; using $B_{0}^{HSM}$ increases $g_{a}(\beta)$ slightly, however, its effect for
a mean free path of $\sim$200 nm remains within the experimental error bar. For this reason and for clarity, we have
used $B_{0}$ which does not depend on the transport mean free path.  We have also deduced the value of $g_{a}(\beta)$ at 4.8 bars from
the magnetic  suppression of the aerogel AB-transition measured by the Lancaster group \cite{Pickett01}, denoted in Fig.\ref{gbeta}
with a star. The solid line is a guide-to-the-eye and the dotted, dashed, and dot-dashed lines are calculations using the homogeneous
scattering model (HSM) which we discuss below. The pressure dependence of
$g_{a}(\beta)$ is  markedly different from the pure superfluid,
$g_{0}(\beta)$, which diverges at the PCP near 21 bars.  The almost linear dependence on pressure observed in the aerogel
superfluid, even at high pressure, is a substantial modification of the phase diagram induced by impurity scattering.

The simplest model of impurity scattering is the Homogeneous Scattering Model (HSM)\cite{Thune98}. In this model, the
scattering probability is independent of position and the medium is completely isotropic. This model has the advantage that the
 Ginzburg-Landau theory is only slightly modified from that of pure $^{3}$He. The superfluid transition in aerogel,
$T_{ca}$, is given by solving Eq.\ref{AGformula} in the form,

\begin{equation} ln(T_{ca}/T_{c0})+\sum_{n=1}^{\infty}\left(\frac{1}{n-\frac{1}{2}}-\frac{1}{n-\frac{1}{2}+x}\right)=0
\label{AG}
\end{equation}where $x\equiv\hbar v_{f}/4\pi k_{B}T_{ca}l_{tr}$ and $l_{tr}$ is the transport mean free path.  The
$\beta_{i}$'s have been calculated for the HSM and are given by\cite{Thune98},

\begin{equation}
\pmatrix{\beta_{1} \cr \beta_{2} \cr \beta_{3}\cr \beta_{4}\cr \beta_{5}} =a_{1}\pmatrix{ -1\cr 2\cr 2\cr 2\cr -2} +a_{2}
\pmatrix{ 0\cr 1\cr 0\cr 1\cr -1} +
\pmatrix{
\delta\beta_{1}^{sc}\cr
\delta\beta_{2}^{sc}\cr
\delta\beta_{3}^{sc}\cr
\delta\beta_{4}^{sc}\cr
\delta\beta_{5}^{sc}}
\end{equation}  where the coefficients $a_{1}$ and $a_{2}$ are given by,
\begin{eqnarray} a_{1}\equiv\beta_{a}^{wc}=\frac{N(0)\sum_{n=1}^{\infty}(n-\frac{1}{2}+x)^{-3}}{240(\pi k_{B}T_{ca})^2},
\end{eqnarray}
\begin{eqnarray} a_{2}=\frac{N(0)\hbar
v_{f}}{288(\pi k_{B}T_{ca})^{3}l_{tr}}(sin^{2}\delta_{0}-\frac{1}{2})\sum_{n=1}^{\infty}(n-\frac{1}{2}+x)^{-4}
\end{eqnarray} and the $\delta\beta^{sc}_{i}$'s are the strong-coupling corrections to the free energies. Note that in the bulk
limit  ($l_{tr}\rightarrow\infty$),  
$a_{1}\rightarrow \beta_{0}^{wc}$ and $a_{2}\rightarrow 0$. We choose a random scattering phase shift $\delta_{0}$ such that
$sin^{2}\delta_{0}=1/2$ and $a_{2}=0$.  Calculations performed with a scattering phase shift in the unitary limit
($sin^{2}\delta_{0}=1$) or  Born limit ($sin^{2}\delta_{0}=0$) have only a small effect on the magnitude of $g_{a}(\beta)$ and do not
alter our conclusions. The strong-coupling corrections relative to the weak-coupling value $\delta\beta_{i}^{sc}/\beta_{0}^{wc}$,
essential to calculate accurately
$g_{a}(\beta)$, are taken from pure
$^{3}$He\cite{Tom}; namely, $\beta_{345}$ is derived from measurements of the B-phase NMR g-shifts and
longitudinal  resonance frequency\cite{Tom}, $\beta_{12}$ from the normal to B-phase heat capacity jump \cite{Greywall}, and 
$\beta_{245}$ and
$\beta_{5}$ are from measurements  of the $A_{1}-A_{2}$ splitting \cite{Israel} and magnetic suppression of the bulk 
AB-transition,
$g_{0}(\beta)$ \cite{Hahn95}. The only unknown parameter in the HSM is the transport mean path, 
$l_{tr}$, for which high-frequency acoustic measurements on our sample were found to be consistent with 
a mean free path of $\sim$200-300 nm \cite{Nomura00,Gervais00}. In
all of the calculations, $T_{ca}$ was solved using Eq. \ref{AG} and was used consistently throughout our 
calculations of the $\beta_{i}$'s.

In Fig.\ref{gbeta}, the dotted line shows the calculations of $g_{a}(\beta)$ which assumes that the strong-coupling corrections
are the same as in the pure superfluid and with a transport mean free path of 200 nm. The effect of scattering is to increase the
polycritical pressure, where
$g_{a}(\beta)$ diverges, thus increasing the stability of the dirty B-phase.  This model does not describe our experimental
data for any reasonable  value of the transport mean free path.  However, we expect that the strong-coupling
corrections should scale to first-order as $\delta\beta_{i}^{sc}\sim (T_{c}/T_{F})\beta_{0}^{wc}$, hence being reduced (relative
to the weak-coupling value) in the dirty
system accordingly to 
$(\delta\beta_{i}^{sc})_{a}/\beta_{a}^{wc}\simeq(\delta\beta_{i}^{sc})_{0}/\beta_{0}^{wc}\times\frac{T_{ca}}{T_{c0}}$
where $(\delta\beta_{i}^{sc})_{a}$ and $(\delta\beta_{i}^{sc})_{0}$ are the  strong-coupling corrections in the dirty and the pure
superfluids. In Fig.\ref{gbeta} the calculations of $g_{a}(\beta)$ were also performed with the HSM using the
rescaled strong-coupling coefficients.  They are shown with two choices of
 transport mean free path of 200 nm (dashed line) and 150 nm (dot-dashed line). Considering  the limitations of the HSM, 
the agreement between the data and calculated
$g_{a}(\beta)$ is reasonably good. However, the better agreemnt with the smaller mean free path of 150 nm
indicates that smaller values of $T_{ca}$ than the epexerimental ones are needed to correctly decsribe the 
strong-coupling corrections in aerogel.  
But more importantly, the HSM calculations shows
unambiguously that the  strong-coupling corrections (relative to weak-coupling) are reduced by impurity scattering and that the
 PCP is increased in a 98\% porous aerogel above the melting pressure, beyond experimental reach.  At the PCP, the heat capacity
jump in the A-phase equals that of the B-phase,  equivalent to the condition, $3\beta_{13}=2\beta_{345}$ in Eq. \ref{gtheory}.
The HSM with rescaled strong-coupling corrections predicts a PCP of $\sim$ 34 bars  for a mean free path of 200 nm and a PCP
of $\sim$ 40 bars for a mean free path of 150 nm. The almost linear  dependence on pressure of
$g_{a}(\beta)$ that we observe suggests that the true PCP may even be higher than estimated from the HSM. Experiments
with higher porosity aerogels, having a correspondingly larger mean free path, may be able to clarify this situation. 
It might
also be necessary to take into account modification of the strong-coupling corrections  beyond a simple rescaling as we have
done, {\it e.g.} the effect of impurities on spin-fluctuation feedback\cite{Georgia}.

The  qualitative agreement of our data with the HSM calculations using a rescaling of the strong coupling corrections to the
$\beta_{i}$'s provides qualitative evidence that the orbital symmetry of the order parameter in the aerogel system is similar to
that of the pure superfluid.  The expression, Eq. 4, for suppression of the AB-transition is specific to the transition between
axial and isotropic states and so the agreement between the model and our measurements of the transition is consistent with, but not a
proof of, their identification with A and B phases in the dirty superfluid.

Finally, it is worth noting that in the pure superfluid  $g_{0}(\beta)$ does not recover its weak-coupling value (1) at zero
pressure, but rather takes the value 1.61 \cite{Hahn95}. This rather large deviation from weak-coupling theory at zero
pressure, $\sim40\%$, is unexpected since all other thermodynamic measurements indicate much smaller deviations from weak
coupling at low pressure. At zero pressure, the B-phase heat capacity jump\cite{Greywall} and $A_{1}-A_{2}$ linear field
splitting
\cite{Israel} are within  3\% of their weak-coupling values. This finite contribution to 
$g_{0}(\beta)$ at zero pressure has been interpreted as evidence that $^{3}$He is not a weak-coupling superfluid at zero
pressure\cite{Hahn95}. As we have shown above, the strong-coupling  corrections are reduced by scattering (scaling with
$T_{ca}$) and should therefore be dramatically reduced at low pressure where $T_{ca}$ is most strongly suppressed. We therefore
expect
$g_{a}(\beta)$ to recover its weak-coupling limit in aerogel at low pressure, if indeed the deviation of
$g_{0}(\beta)$  from weak coupling in pure $^{3}$He at low pressure is due to strong-coupling corrections. Our data in aerogel do not
allow us to extrapolate to zero pressure and so an accurate measurement of $g_{a}(\beta)$ at low pressures would be desirable since it
might shed light on this problem.

\subsection{Field dependence of $T_{ca}$}

\begin{figure}[t]
  \begin{center}
   \leavevmode
    \hbox{\epsfxsize=.9\columnwidth{\epsfbox{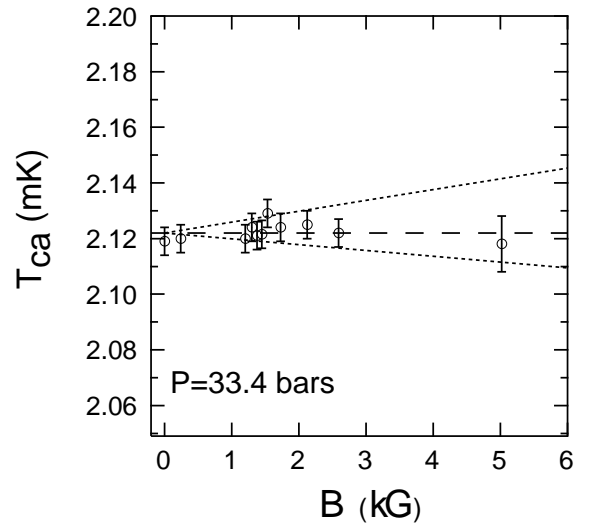}}}
  \end{center}
  \caption{Field dependence the aerogel superfluid transition $T_{ca}$ at 33.4 bars. The dashed line shows the average value of
$T_{ca}$. The slopes of the two dotted lines are the same as for the $A_{1}-A_{2}$ splitting measured for pure $^{3}$He near
melting pressure
\protect\cite{Israel}.}
  \label{tcfield}
\end{figure}

The field dependence of $T_{ca}$ at 33.4 bars is shown in Fig.\ref{tcfield} for fields ranging from 0 to 5 kG. The transition is
clearly field independent and from this we can infer that  the superfluid transition is from a normal fluid to an ESP
superfluid. These data contradict earlier NMR work where a $B^{2}$ field dependence was reported for the superfluid transition
in aerogel\cite{Spra96}. Upon reanalysis of these same earlier results including some additional experiments, by Haard\cite{Tom}, 
it was found that the NMR data  are consistent with the acoustic experiments presented here to within experimental error.

In pure superfluid $^{3}$He, owing to  particle-hole asymmetry, a magnetic field produces a linear splitting of the normal to
A-phase transition 
\cite{Israel}. The field dependence of the  splitting between the $A_{1}$ and $A_{2}$ phase lines is 6
$\mu$K/kG near melting pressure and is shown in Fig.\ref{tcfield} as dotted lines shifted to the average value of $T_{ca}$.

In contrast, for the aerogel superfluid we do not observe any deviation of $T_{ca}$ from its zero-field value, nor do we
observe significant broadening of the transition region in the acoustic trace even at our highest field
 of 8 kG (not shown in Fig.\ref{tcfield}. This splitting may be hard to resolve owing to the rather broad superfluid transition in
aerogel, $\sim 30\mu K$. The expected linear  $A_{1}-A_{2}$ splitting in the dirty superfluid can be estimated from calculations based
on Ginzburg-Landau theory  and the HSM (see above). The bulk
$A_{1}-A_{2}$ splitting can be expressed by the quantity $U_{0}$ defined
 as,\cite{Hahn95}

\begin{equation} U_{0}\equiv -\frac{(dT/dB)_{A1}}{(dT/dB)_{A2}}=-\frac{\beta_{5}}{\beta_{245}}.
\end{equation} which ranges for pure $^{3}$He from 0.97 at zero pressure to 1.81 at melting pressure.  In the HSM, the $U$ 
parameter can be calculated from  $\beta_{5}$ and $\beta_{245}$ and  in aerogel takes the value at melting pressure of
$U_{a}\simeq 0.85 U_{0}$ with a mean free path of 200 nm.   For this estimate, we have used the rescaled  bulk strong-coupling corrections
as discussed above.  Measurements of $T_{ca}$ in high fields sufficient to resolve this splitting  are desirable in order to
establish a better understanding of the thermodynamics of the dirty superfluid.

\subsection{Supercooling of the aerogel AB-transition in zero applied field}

We have shown that the polycritical point vanishes for superfluid $^{3}$He in 98\% aerogel.  Consequently,
in zero field we expect that superfluidity occurs by a second order transition from the normal state directly to the B-phase
without supercooling.   Our observation of supercooling giving a large region of metastable A-phase {\it with no
applied field} is quite unexpected.  Supercooling was  first noted by Barker {\it et al.}\cite{Barker00} for the aerogel
AB-transition at 284 G with
$^{4}$He preplating.  This was also found for zero applied field and was studied as a function of magnetic field by Gervais {\it
et al.}\cite{Gervais01,Gervais02}.  Our smallest field in this case was estimated
to be less than 10 G. There are a number of possible intrepretations and so a review of the expected behavior in the limit of
low magnetic field is appropriate.   

Taking pure superfluid
$^{3}$He as a guide, the first transition encountered on cooling in a small but non-zero magnetic field is from the normal
state to the
$A_{1}$-phase and then, in the absence of supercooling, to the B-phase. If there were to be supercooling it is likely that the
$A_{2}$ transition would be encountered before reaching the B-phase.  This follows since the $A_{1}$ transition increases
linearly with field whereas the B-phase is suppressed quadratically.  Consequently, the supercooling of the aerogel
AB-transition at 284 G observed by Barker {\it et al.}
\cite{Barker00}  can be explained in a trivial way: a small region of  $A_{1}$-phase in equilibrium, $\alt$ 1$\mu$K, is first
encountered on cooling, leading to a metastable A-phase that supercools until the B-phase nucleates. However, the mechanism for
producing relative stability of one phase over another, or the nucleation of the most stable phase, has not been explored in
such a small interval of temperature near $T_{c}$.

\begin{figure}[t]
  \begin{center}
   \leavevmode
    \hbox{\epsfxsize=1.\columnwidth{\epsfbox{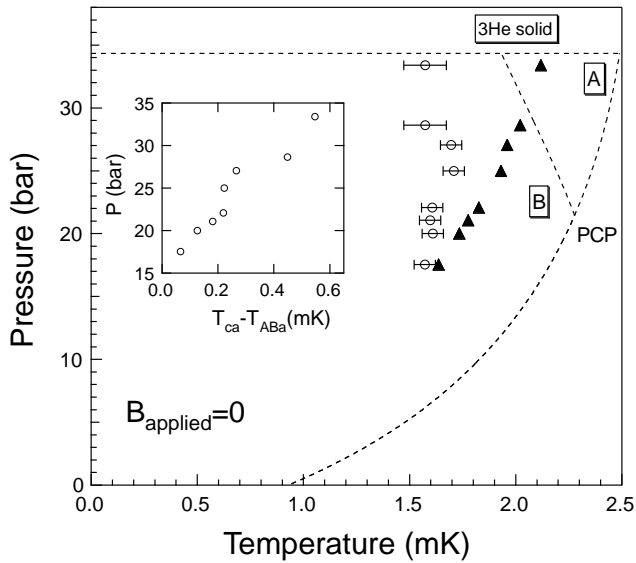}}}
  \end{center}
  \caption{Phase diagram for metastable $^{3}$He A-phase in 98\% porous aerogel in zero applied magnetic field. The triangles
are the aerogel superfluid transition and the open circles the aerogel A-B transition on cooling.  The inset shows the
magnitude of supercooling of the aerogel A-B transition as a function of pressure. The dotted lines are the pure $^{3}$He
superfluid phase with  A, B and solid phases indicated.}
  \label{diagram}
\end{figure}

Can the same argument for supercooling in 284 G  hold at very low field as well? If the total field for the case of
zero applied field were less than 10 G then the window of stability of the $A_{1}$-phase would be only $\alt$ 0.03 $\mu$K.  We
find supercooling under these circumstance to be even more remarkable since there is no evidence of similar supercooling for
the pure superfluid at pressures below the polycritical point.  If the phenomenon of supercooling in low field  were to be
unique to the dirty superfluid then it would require a correspondingly unique nucleation scenario specific to the aerogel
AB-transition, which, in addition,  must also account for our observation that the extent of supercooling
is field independent\cite{Gervais01}.   Furthermore, it seems not to make sense to rely on details for phase stability in
such a narrow window of temperature when the superfluid transition
$T_{ca}$ is inhomogeneously broadened over an interval three orders of magnitude larger.  A second possibility, and one that we
believe to be more likely,  is that there must be a thin, but unobserved, sliver of A-phase near
$T_{ca}$ which gives rise to the metastability that is observed. This sliver  is at most
$\alt$20
$\mu$K wide, but might be stabilized by inhomogeneity in the aerogel structure in a manner that is not yet understood. In
Fig.\ref{diagram}, the metastable phase diagram from Gervais {\it et al.}\cite{Gervais01} is shown with no applied
magnetic field ranging in pressure from 15 to 33.4 bars.  The region of metastability is strikingly similar to that observed by
Barker {\it et al.}\cite{Barker00}  in a field of 284 G, with
$^{4}$He preplating, and for a much larger aerogel sample.  The critical radius $R_{c}$ for B-phase nucleation may play an
important role, since a large critical radius in the aerogel might  give rise to the observed metastability. An estimate\cite{Gervais02}
near melting pressures from the susceptibility difference and field dependence of $T_{ABa}$ shows that it is slightly larger in
aerogel, $R_{c}^{aero}\sim 5R_{c}^{bulk}$, at the same value of $T/T_{AB}$. In pure $^{3}$He the
critical radius has not been measured below melting pressure. Future work in a magnetically shielded environment and with
different aerogel densities,  together with characterization of
$R_{c}$  as a function of pressure,  could bring some understanding to this puzzle. 

For all experiments to date it has been unavoidable that pure and dirty superfluids be juxtaposed.  In future work it may be possible
to explore the connection between these two unconventional superfluids and such experiments can benefit from detailed knowledge of
the phase diagram.  In this spirit we have recently investigated the nucleation of the AB-transition\cite{Gervais02} showing that
proximity coupling between the pure and dirty superfluids is too weak to act as a source of nucleation.

\section{Conclusion} We have described the phase diagram of superfluid $^{3}$He in 98\% aerogel. In this phase diagram we find
two equilibrium states which we call A and B, by analogy with pure $^{3}$He.  The B-phase is favored in zero field and
is destablized by a magnetic field yielding the A-phase.  The superfluid transition from the normal state, {\it i.e.} normal to
A-phase transition, is insensitive to magnetic field.  On this basis alone we can be confident of the nature of the spin part
of the order parameter associated with each of these phases: the A-phase is an equal spin pairing state and the B-phase is a
non-equal spin pairing state.  The field dependence of the AB-transition can be understood from calculations using the HSM
model combined with a simple rescaling of
 strong-coupling corrections to the quasiparticle interactions, assuming that the A and B dirty phases are in fact the axial and
isotropic {\it p}-wave states. The theory and  experiment both concur that sufficient impurity scattering, as is the case for 98\%
aerogel, causes the polycritical point to vanish.  It is intriguing that no hint of the expected but small $A_{1}-A_{2}$ splitting was
observed even at fields of
$\sim$8 kG; however, higher field experiments will be better able to address this question. 
 
The extensive supercooling of the AB-transition, especially with no applied field, remains  a puzzle.  It cannot be simply
explained in terms of the phase diagram that we present, nor in terms of current nucleation scenarios.  This raises further 
questions concerning the nature of the superfluid state in aerogel
and whether it might be inherently inhomogeneous.

In summary, the measurements of the equilibrium phase diagram of $^{3}$He with impurity scattering improves our understanding of
pure and dirty superfluids  and sets the stage for a better understanding of its non-equilibrium behavior, metastability and
nucleation.

\acknowledgments{

We acknowledge helpful discussions with J.A. Sauls, E.V. Thuneberg and G. Kharadze. Work supported by the NSF under grant no.
DMR-0072350.  One of the author (G.G) acknowledges receipt of support from FCAR (Qu\'ebec).}

\end{document}